# Resistance of the cell wall to degradation is a critical parameter for isolation of high quality RNA from natural isolates of *Bacillus subtilis*

Jean-Sébastien Guez[a], François Coutte[a], Anne-Sophie Drucbert[b], Pierre-Marie Danzé[b] and Philippe Jacques[a]

[a] Laboratoire ProBioGEM, UPRES-EA 1026, Polytech-Lille, IUT A, Université des Sciences et Technologies de Lille, Bd Paul Langevin, F-59655 Villeneuve d'Ascq

[b] IFR114-IMPRT (functional genomic platform), Faculté de Médecine H. Warembourg, Place de Verdun, F-59045 Lille

**Abstract**

Natural isolates of *Bacillus subtilis* are known for their ability to produce a large panel of bioactive compounds. Unfortunately, their recalcitrance to conventional molecular techniques limits their transcript studies. In this work, difficulties to isolate RNA attributed to the cell wall were overcome, finally authorizing powerful RT-PCR's

**Keywords**

**Manuscript**

The endospore-forming bacterium *B. subtilis* ATCC 6633 is a natural isolate with a high potential of peptide antibiotic production, whether of ribosomal origin, as subtilin (Leenders et al. 1999) and subtilosin (Stein et al. 2004), or of non-ribosomal origin, as rhizocticin, mycosubtilin and surfactin (Duitman et al. 1999; Leclère et al. 2005). Most of these peptides are not produced by the reference laboratory strain *B. subtilis* 168 (Zeigler et al. 2008). Regulation studies of the biosynthesis of such compounds have to be made in *B. subtilis* ATCC6633. A counterpart is the recalcitrance of this isolate to molecular techniques like genetic transformations (Duitman et al 2007; Leclère et al. 2005) or transcript analysis requiring hence RNA extraction.

Owing to the different characteristics of the cell wall structure and composition, the isolation of high quality RNA from gram-positive organisms is known to be more difficult than for gram-negative ones. There are many methods for disrupting gram-positive cell walls for RNA isolation purposes. These methods can act enzymatically through a lysozyme and proteinase K treatment or chemically with homogenisation of the cells in a lysis buffer (Van Dessel et al. 2004). They can also act mechanically through the use of violently agitated zirconia-silica beads (Hambraeus et al. 2003; Moeller et al. 2006) or glass beads (Putzer et al. 1992; Oh and So 2003). In a preliminary work (data not shown), such routine techniques for cell lysis in the RNA isolation protocol were efficient for *B. subtilis* 168 Marburg but not for the natural strain *B. subtilis* ATCC 6633.

To analyse the cell wall resistance to degradation, the kinetics of protoplast generation of both strains were compared. They were grown in Erlenmeyer flasks at 30 °C and 160 rpm in a modified Landy medium (Guez et al. 2007) buffered with MOPS (3-(*N*-morpholino)propanesulfonic acid, Sigma-Aldrich, France) 100 mM, pH 7.0, supplemented with tryptophan, 16 mg/l. Before inoculation, the precultures were centrifuged and washed in 9 g/l NaCl solution. The initial optical density of the culture was between 0.05 and 0.1 OD units, measured at 600 nm using an Uvicon 922 spectrophotometer (Kontron, Germany). Protoplast generation kinetics were obtained using a previously described protocol (Errington 1990). For cells sampled at 1 OD unit, corresponding to the early exponential growth, complete protoplast formation was obtained with *B. subtilis* 168 after 30 min of incubation in the SMPP buffer added with 2 mg/ml lysozyme, whereas a partial protoplast formation rate was obtained with *B. subtilis* ATCC 6633, 83 % after 30 min and 93 % after 60 min of incubation (Table 1). For cells sampled at 5 OD unit, corresponding to the middle-late exponential growth, a higher partial protoplast formation rate was obtained with *B. subtilis* 168, 70 % after 30 min and 82 % after 60 min, whereas a more partial one was obtained with *B. subtilis* ATCC 6633, 43 % after 30 min and 59 % after 60 min. This experiment showed the lower protoplast formation rates obtained with *B. subtilis* ATCC 6633 compared to *B. subtilis* 168.

Taking into account this information, some modifications of the Ribopure (Ambion, Germany) procedure for RNA isolation from *B. subtilis* ATCC 6633 were experimented. A volume corresponding to $5 \times 10^8$ cells was sampled on cultures and immediately added to the RNALater ice-cold solution (v/v) for rapid RNase inactivation, mixed during 10 s and centrifuged 5 min at −9 °C and $11,000 \times g$. The supernatant was discarded and the pellet stored at −80 °C. The defrozen cells were then resuspended in 50 μl of a buffer solution and treated with lysozyme (table 2) (10 mg/ml, Sigma-Aldrich, France). Different incubation conditions in the lysozyme solution were tested, 0, 10 or 30 min, at room temperature. It should be noted that incubation with lysozyme at 37 °C was not recommended because of extensive shearing of RNA. A volume of 350 μl of RNAWiz phenolic solution (RiboPure Bacteria, Ambion, Germany) was added and a mechanical cell lysis in the presence of 250 μl of zirconia beads was performed for 0, 10 or 30 min at high speed using a vortex adaptater.

The lysate was then centrifuged 5 min at 12,000 × *g* and 4 °C. The beads were discarded and 0.2 volume of chloroform added. The mixture was shaken for 30 s, incubated 10 min at room temperature and centrifuged 5 min at 12,000 × *g* and 4 °C. The aqueous phase was then retrieved and 0.5 volume of ethanol added. Purification of RNA was completed on glass-fiber using the RiboPure protocol. An additional DNase treatment was performed adding 5.5 µl of 10 X DNase buffer (Ambion) and 4 µl of DNase (Ambion) followed by 30 min incubation at 37 °C. The solution was then mixed with 10 µl of the inactivation reagent, incubated 2 min at room temperature, centrifuged 1 min at 12,000 × *g* and the supernatant was stored in a tube. The total quantity of RNA was measured with a Nanodrop ND-1000 spectrophotometer (Agilent, USA). The RNA 23S/16S ratio (RNA 6000 NanoAssay, 2100 Bioanalyzer, Agilent, USA) and the electropherograms were used as a quality indicator of RNA. Electrophoresis were performed in non-denaturing conditions (Jahn et al., 2008). As the use of the RIN (RNA integrity number) can not be formally performed for prokaryotes yet (Schroeder et al. 2007), the electropherograms were analysed with giving importance to the baselines of the fast region and of the ribosomal inter-region which are critical for assessing quality of RNA. The protein contamination was estimated by measuring the $A_{260\,nm}/A_{280\,nm}$ ratio (Nanodrop ND-1000).

Compared to the very low RNA extraction yield of 0.3 µg obtained in the absence of lysozyme, its presence allowed, in some cases, a better RNA extraction (Table 2). In the absence of incubation, the low yields of 2.3 and 4.5 µg were obtained because of insufficient cell wall degradation. Nevertheless, the value of the 23S/16S ratio between 1.83 and 1.98 showed the good quality of the RNA. With an incubation time of 30 min in lysozyme, the low RNA yields of 3.4 and 5.7 µg were probably obtained because of extensive RNA degradation which is confirmed by the low 23S/16S ratio of 1.04 and 1. With an incubation time of 10 min in lysozyme, RNA extraction yielded 13.5 and 15.0 µg. For this latter condition, the quality of RNA was better when agitating for 10 min in the presence of zirconia beads, 23S/16S ratio of 1.71, than for 30 min, 23S/16S ratio of 1.24. The gels presented the absence of smears (data not shown) and electropherograms showed the absence of deviation of the baseline, especially for the segments of the fast region and the ribosomal inter-region. According to the criteria developed previously by authors (Schroeder et al. 2006), the extracted RNA would belong to the RNA integrity categories 7 or 8, corresponding to minor RNA degradation. It is to noticed that RIN values above 7 are considered as sufficient values for further transcript analysis (Jahn et al. 2008).

With such a protocol, comparative RNA isolations were then performed on culture samples from *B. subtilis* ATCC 6633 and 168 taken after 4, 6 and 7 generations which corresponded to the middle and late exponential growth phase (figure 1). The RNA specific extraction yields expressed in µg of RNA per $10^8$ cells were lower for *B. subtilis* ATCC 6633, ranging from 1.8 to 0.11 according to the sampling time, compared to *B. subtilis* 168, ranging from 2.4 to 0.23. The lowest yields for both strains corresponded to samples taken at 24 h, *i.e.* during the stringent response, which is a well known phenomenon. Interestingly, the evolution of these yields for increasing generation numbers also showed a greater drop for *B. subtilis* ATCC 6633, observed after 6 and 7 generations, compared to *B. subtilis* 168. These results globally showed the efficiency of the RNA isolation method for early-growing cells and the increased difficulty in isolating RNA from late-growing cells, independently from the ecological origin of the cells. They particularly confirm the recalcitrance of *B. subtilis* ATCC 6633 to molecular techniques (Duitman et al 2007) which is caused by the cell wall being subjected to different dynamic structural and biochemical modifications.

A transcript analysis was finally performed to confirm the high quality of RNAs prepared from *B. subtilis* ATCC 6633. The expression of *cspB* gene coding for a major cold shock protein was done on cells sampled on cultures made at different temperatures. The semi-quantitative measurement of *cspB* transcripts was done by RT-PCR following the method presented previously by authors (Guez et al. 2008). RT-PCR were done with 3 µg of prepared RNAs. To estimate the level of expression of the *csp* gene coding for the major cold shock protein, a 154 bp PCR product belonging to the *csp* locus was amplified using the following pair of oligonucleotides: 5'-AAAAGGTTTCGGATTCATCG-3' and 5'-AACGTTAGCAGCTTGTGGTC-3'. It was verified that the expression of *rplL*, a housekeeping gene of *B. subtilis* coding for the ribosomal protein L12, remained constant for each sample. RT-PCR results of *cspB* gene expression for cultures made at 20, 30 and 37°C are shown on figure 2. As the level of expression of *cspB* increased drastically with decreasing the temperature of the cultures, which was the expected physiological response (Willimsky et al. 1992), this RNA isolation method was found to be particularly well suited for an example of transcript analysis and could be very useful for further work enabling larger gene expression studies of natural isolates of *B. subtilis*.

**Acknowledgements**

This work received the financial support from the Université des Sciences et Technologies de Lille (BQR), the Agence Nationale de la Recherche (ANR) and the European Funds for the Regional Development. Authors wanted to thank Dr W. Everett for kind re-reading.

**Figures and tables**

Table 1 : Evolution of the protoplast formation rate expressed in % after 30 and 60 min of incubation in the SMPP buffer added with 2 mg/ml of lysozyme. Samples were taken at the stage of early and middle-late exponential growth during cultures of *B. subtilis* 168 and ATCC 6633 in the modified Landy medium. Standard deviations were calculated at least on a duplicate set of cultures.

Table 2 : RNA isolation yield (μg), purity analysis ($A_{260\ nm}/A_{280\ nm}$ ratio), and quality indicator (electropherograms, 23S/16S ratio,) obtained with *B. subtilis* ATCC 6633 ($5\times10^8$ cells/ml) for different lysozyme incubation conditions (0, 10 and 30 min) and for different agitation times (0, 10 and 30 min) in the presence of zirconia beads. Lysozyme at 10 mg/ml was added prior to the RNA isolation protocol. Standard deviations were calculated on a duplicate set of analysis. Electropherograms are representative of a duplicate set of analysis.

Figure 1 : RNA specific extraction yields (μg of RNA per $10^8$ cells) from *B. subtilis* ATCC 6633 and 168 cultivated in the modified Landy medium for a number of generations of 4, 6 or 7 or for a time of 24 h. Standard deviations were calculated on a triplicate set of cultures.

Figure 2 : Comparative RT-PCR gel of *rplL* and *cspB* genes from *B. subtilis* ATCC 6633. Samples were prepared from growths obtained in flasks at different temperatures 20, 30 and 37°C. The culture medium was the modified Landy medium buffered with MOPS 100 mM at pH 7.0.

Table 1

|  | *B. subtilis* 168 | | *B. subtilis* ATCC 6633 | |
| --- | --- | --- | --- | --- |
|  | Early growth | Middle-late growth | Early growth | Middle-late growth |
| Protoplast formation rate after 30 min (%) | >99 | 70 ± 1 | 83 ± 4 | 43 ± 4 |
| Protoplast formation rate after 60 min (%) | >99 | 82 ± 3 | 93 ± 4 | 59 ± 1 |

Table 2

| Experimental conditions | | RNA yield (µg) | $A_{260}/A_{280}$ ratio | Electropherogram | 23S/16S ratio |
|---|---|---|---|---|---|
| Incubation time in lysozyme | Agitation time in the presence of zirconia beads | | | | |
| No incubation | No agitation | < 0.1 | / | / | / |
| | 10 min | 2.3±1.2 | 2.07±0.03 | 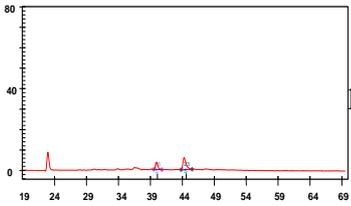 | 1.98±0.15 |
| | 30 min | 4.5±1.5 | 2.08±0.02 | 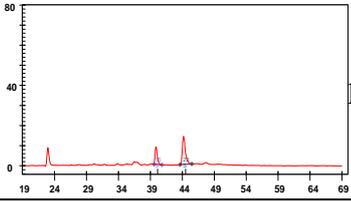 | 1.83±0.12 |
| 10 min | No agitation | < 0.1 | / | / | / |
| | 10 min | 15.0±0.9 | 2.13±0.02 | 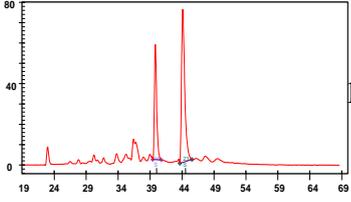 | 1.71±0.06 |
| | 30 min | 13.5±0.2 | 2.14±0.01 | 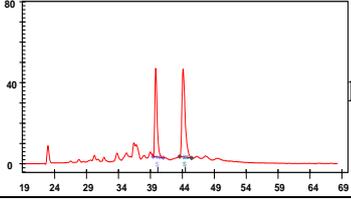 | 1.24±0.11 |
| 30 min | No agitation | < 0.1 | / | / | / |
| | 10 min | 3.4±0.8 | 2.11±0.03 | 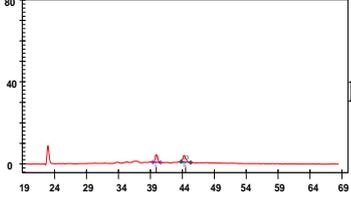 | 1.04±0.20 |
| | 30 min | 5.7±1.3 | 2.12±0.01 | 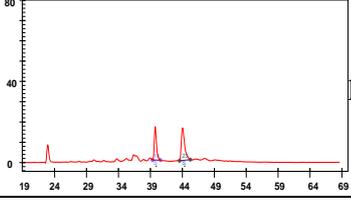 | 1.32±0.08 |
| No lysozyme added [*] | 10 min | 0.3±0.1 | / | / | / |

(*): Ambion Ribopure reference protocol